# Towards Digital Engineering

# -- The Advent of Digital Systems Engineering

Jingwei Huang*, Adrian Gheorghe, Holly Handley,
Pilar Pazos, Ariel Pinto, Samuel Kovacic, Andrew Collins,
Charles Keating, Andres Sousa-Poza, Ghaith Rabadi,
Resit Unal, Teddy Cotter, Rafael Landaeta, Charles Daniels

Department of Engineering Management and Systems Engineering
Old Dominion University
2101 Engineering Systems Building, Norfolk, VA 23529, USA
Email: j2huang@odu.edu
*Corresponding author

**Abstract:** Digital Engineering, the digital transformation of engineering to leverage digital technologies, is coming globally. This paper explores digital systems engineering, which aims at developing theory, methods, models, and tools to support digital engineering practice. A critical task is to digitalize engineering artifacts, thus enabling information & model sharing, traceability, and accountability across platforms, across lifecycle, and across domains. We identify challenges and enabling digital technologies; analyse the transition from traditional engineering to digital engineering; define core concepts, including *digitalization*, *unique identification*, *digitalized artifacts*, *digital augmentation*, and others; present a big picture of digital systems engineering in four levels: vision, strategy, action, and foundation; briefly discuss each of main areas of research. Digitalization enables fast infusing and leveraging novel digital technologies; unique identification enables information traceability and accountability in lifecycle; provenance enables tracing dependency relations among engineering artifacts, supporting model reproducibility and replicability, and helping with trustworthiness evaluation of digital artifacts.

## 1. Introduction

In order to rapidly infuse innovative digital technologies and to meet the new demands from the digitalizing world, Digital Engineering, the digital transformation of engineering, is emerging with different names globally, such as Industry 4.0 (GTAI, 2014), digital manufacturing or smart manufacturing (White House, 2012, 2018), and others. Many problems appeared in traditional engineering and acquisition processes, such as linear engineering process to develop complex systems, document-intensive and stove-piped information flow, hard to change and sustain systems in rapidly changing and uncertain operational environment, and others. To address those problems, the US Department of Defense (DoD) launched the Digital Engineering Strategy (DES) (US DoD, 2018), aiming to build digital enterprise and to fast incorporate technological innovation by means of digitally representing the system of interest, developing, using, integrating and curating formal model across organizational boundaries and lifecycle activities, and using "Authoritative Source of Truth" as the central platform and repository for collaborating, communicating, and sharing data and models. This strategy exhibits a profound vision for "transforming engineering practices to digital engineering and incorporating technological innovations to produce an integrated, digital, model-based approach" (US DoD, 2018). Towards this direction, research efforts combining digital technologies into systems engineering (SE) to meet the new demands from the digitalizing world are converging into an emerging field – digital systems engineering, which aims at developing knowledge and technology to support digital engineering practice. This paper explores this exciting area.

First of all, let us briefly look into the new landscape of engineering systems. The fast-growing Internet of Things (IoT) (Atzori, Iera, & Morabito, 2010; IEEE, 2015) is dramatically changing the world; IoT has become a trigger for numerous innovative applications, leading to various cyber-physical-social smart systems (CPS3) (Huang, Seck, & Gheorghe, 2016). Particularly, Industrial IoT (IIoT) (Sisinni, Saifullah, Han, Jennehag, & Gidlund, 2018; H. Xu, Yu, Griffith, & Golmie, 2018) and industrial smart CPSs are paving the way to the fourth industrial revolution (Schwab, 2017), or in short, so-called Industry 4.0 (GTAI, 2014; L. Da Xu, Xu, & Li, 2018). IoT is quickly changing the landscape of engineering systems from the beginning of systems design through the end of lifecycle. By using IoT, Big Data technologies, AI, and Machine Learning (ML), the fingerprints (or footprints) of a system (i.e., the dynamic changes of system status and changes of components and behavior) can be potentially traced in the whole system lifecycle. Similarly, the dynamic changes of the system's operating environment can be observed, recorded, and mined to provide valuable information for engineering systems design, testing, manufacturing, operations, maintenance and support, reuse and recycle, as well as risk analysis toward trustworthy and resilient systems.

Highly associated with IoT, digital transformation, a term reflecting the pervasive diffusion of digital technologies in engineering, business, and many societal processes, is profoundly changing almost every aspect of human being's activities, from our daily life to various businesses, including science and engineering.

Jim Grey had a vision that science is transforming into the fourth paradigm – data-intensive paradigm, after empirical, theoretical, and computational (Hey, Tansley, & Tolle, 2009). In the data-intensive paradigm, the essential activities are data capture, data curation, knowledge discovery from data, and data publishing. A theory is an abstraction of the known knowledge about a system, thus having limitations. Data (the observations of a system) can bring in new insight for better understanding and can provide opportunities for new findings and get a breakthrough towards establishing a new theory. This vision has inspired data-intensive research in many science & engineering disciplines and the development of Data Science (Hey et al., 2009; NIST, 2015).

Digital engineering incorporates digital technologies such as IoT, smart cyber-physical systems, big data, AI, ML, robotics, virtual reality (VR), augmented reality (AR), digital twin (Glaessgen & Stargel, 2012; Tao et al., 2018), 3D printing, digital trust, and blockchain (Katina, Keating, Sisti, & Gheorghe, 2019; Nakamoto, 2008; Wang et al., 2019). A remarkable example is the onset of digital manufacturing. Digital engineering is a manifestation of digital transformation in the field of engineering.

The contents of this paper are organized as follows. Section 2 discusses the goals of DoD DES; section 3 discusses the significant challenges to achieving those goals; section 4 briefly discusses critical enabling technologies; then in section 5, we present the framework of digital systems engineering, defines a small set of core concepts, and discusses areas of interest in four levels: vision, strategy, action, and foundation with enabling technologies; finally, section 6 concludes the paper and briefly discuss further research.

## 2. Goals of Digital Engineering Strategy

The central theme of the DoD's Digital Engineering Strategy (US DoD, 2018) is to digitally represent systems of interest and to enable formal model development, integration and use across the system lifecycle phrases through "Authoritative Source of Truth", as illustrated in figure 1. The strategy identifies five tightly related goals.

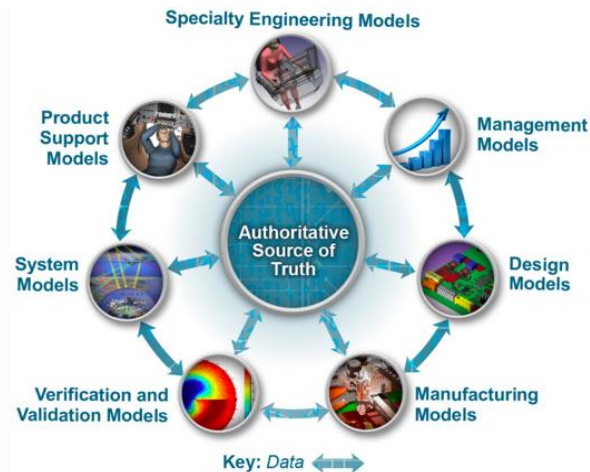

Fig. 1. Illustration of models connected via Authoritative Source of Truth in US DoD Digital Engineering (from (US DoD, 2018), Fig.4).

The first goal is fundamental, which targets the formalized planning of model creation, curation, integration, and use to support decision making, by advocating digitally representing the system of interest; by establishing policy, guidance, rules and standardized syntax, semantics, and lexicons for model development and reuse; by capturing and maintaining model provenance to enable traceability as a basis of judging model trustworthiness for model reuse; and by curating a set of standardized models in the "Authoritative Source of Truth" to enable collaborative engineering activities and decision making across different disciplines and organizations in the system lifecycle.

The second goal targets the establishment of trustworthy knowledge infrastructure, called "Authoritative Source of Truth" (AST), for hosting and sharing across the lifecycle the standardized models, data, and other digital artifacts, which are traditionally isolated within the boundaries of organizations or disciplines. AST supports to capture and curate the history of model evolution through the engineering lifecycle, to maintain the traceability, and propagates the updated models and data to all affected systems and entities for supporting the coordination of associated activities, thus "to enable delivery of the right data to the right person for the right use at the right time." Stakeholder organizations will establish policies and procedures to govern the proper use of AST, including - access control to ensure access by only the authorized users, the use of AST as technical baseline to support engineering decision making on cost, schedule, and performance, to support technical review, and to support communication and collaboration across teams and organizations.

The third goal aims to establish an end-to-end digital enterprise operating in the digitalized and connected environment, to rapidly innovate, infuse, and adopt advanced technologies such as big data analytics, cloud computing, AI & ML, virtual reality, augmented reality, digital twins, digital manufacturing, 3D printing, and many others; and to advance human-machine interactions.

The fourth goal aims to transform the current IT infrastructures and environment, which are "often stove-piped, complex, and difficult to manage, control, secure, and support" (US DoD, 2018), into digital engineering infrastructures and environment, which are expected to be "a more consolidated, collaborative trusted environment" (US DoD, 2018). Digital engineering infrastructures and environments will need able to provide: (a) secure connected information networks supporting computing and information flows at all security levels; (b) the associated evolving digital engineering methods, processes, and tools for visualization, analysis, model management, model interoperability, workflow, collaboration, and extension/customization support; (c) cybersecurity to secure IT infrastructures and to protect intellectual property such as patents, copyrights, trademarks, and other commercial proprietaries through collaborative efforts between government and industrial partners.

Finally, the fifth goal targets the transformation of culture and workforce, including - advance digital engineering policies, standards, and guides; accommodate digital engineering development; digital engineering management; building and preparing workforce via training and education.

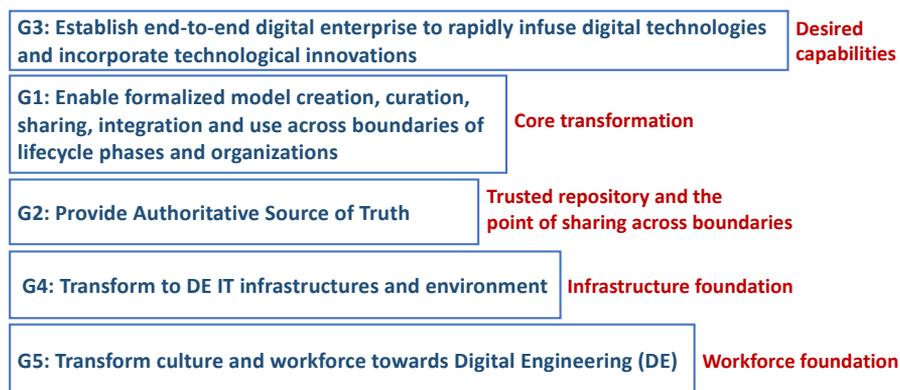

Fig. 2. Relations among Goals of DoD Digital Engineering Strategy

The relations among those goals can be illustrated as a goal stack as shown in figure 2, where from bottom to top,
- Goal 5 is the organizational and human foundation aiming at transforming culture and workforce to provide the eco-environment for the growth of digital engineering, which is fundamental to all other goals;

- Goal 4 transforms traditional IT into new IT infrastructures and service environment for digital engineering. Goal 4 is the IT basis for others;
- Goal 2 creates and maintains the Authoritative Source of Truth, which is the repository and access portal for all standardized models and other digital artifacts, with support from goal 4, and is a basis for goal 1;
- Goal 1 transforms traditional engineering towards formalized model creation, curation, sharing, integration, and use across the boundaries of lifecycle phases, disciplinary teams, and organizations, with support from goal 2;
- Goal 3 establishes an end-to-end digital enterprise to quickly infuse advanced digital technologies and keep rapid innovations, with support from goal 1. Goal 3, as a driving force, produces new requirements for all other goals.

What is in the core of the digital engineering strategy is the digital transformations requiring:

- Digital representation of the system of interest (including not only the focused system and its components, but also possibly the relevant processes, equipment, products, parts, functions, services, and other relevant systems in the operating environment);
- Use of Authoritative Source of Truth (AST) as the repository and the access portal of standardized models, data, and other digital artifacts;
- Formalized model creation, curation, sharing, integration, and use across the boundaries of disciplinary teams, organizations, and the lifecycle phases, with support of AST.

## 3. Challenges

Given the targeted goals, the core digital transformations in need, and the current engineering practice, there are many challenges ahead on the way of digital engineering transformation.

**Challenge 1: Big Data issues in digital engineering** – In the envisioned digital engineering operating in digital and connected environment, every engineering process will face unprecedented Big Data from upstream engineering processes, from engineering partners, from the lifecycle of the system of interest and previous engineering systems (older versions or similar ones), from interacting external systems, from system operating environment, from supply-chain and manufacturing environment, from stakeholders, and others. Those data have not only unprecedented large size but also various forms of different qualities and possible in high velocity of streaming in. The big data brings both opportunities and challenges. On the one hand, an engineering team can leverage new information to improve the quality and to reduce engineering time and cost; on the other hand, the big data poses a significant challenge regarding how to quickly process, manage, mining, analysis, integrate those data and shared models in digital engineering practice. It is also a challenge for engineering teams to collect, manage, and share the data and models produced in their engineering process.

**Challenge 2: Centralized standardization vs. distributed evolutionary standardization** -- Standardized or commonly shared digital representation forms, semantics, and vocabulary are critical for sharing digitalized engineering artifacts (particularly models). In a centralized approach of standardization, a standardized form is defined for digital representation, and the whole community stays with the standard. In a distributed evolutionary approach, ontologies are developed in a crowdsourcing fine-grained evolving process, in which many working groups develop their versions of ontologies by constructing new ontology parts on top of useful parts from other existing ontologies. In this way, some ontology parts, which are most commonly used by a community, naturally converge to a "standard" language. The DoD digital engineering strategy appears towards centralized standardization, which could be a fast and effective engineering approach for the US DoD community. However, there exists a risk of stiffing innovations if standards are applied too early in emerging new fields of technological innovation. Also, business partners worldwide may have their own standards, and the compatibility between standards could be a challenge. It is always a challenge to choose between centralized approaches and distributed evolutionary approaches. Generally speaking, compared with distributed approach, a centralized approach may be more efficient and more effective in the current context, and at least in the near term. However, it also has two related significant issues: a single point of failure and heading in a wrong direction from a long term view. It is of paramount importance to keep and maintain diversity in the course of evolution, no matter to biological populations or technological approaches.

**Challenge 3: Centralized vs. distributed mechanisms of trust** – In the US DoD DES, Authoritative Source of Truth plays a fundamental role in assuring the trustworthiness of models about credibility, accuracy, reusability, safety, and security, as well as other concerned qualities. The challenging issues between centralized or distributed approaches also exist here. The proposed AST appears a centralized mechanism of trust; again, this could be a fast and effective engineering approach. However, given the fact that today's industry has long and complicated supply-chains and many system components could come from allies and trading partners, how to efficiently and effectively incorporate distributed mechanisms of trust in AST is a challenge.

**Challenge 4: Balancing access and control in AST** -- DoD DES aims "to enable delivery of right data to the right person for the right use at the right time" via AST. Access control has been a delicate matter for decades. The challenge still stands in digital engineering. Many entities involved in the engineering workflows may use different access control models and policies, such as MAC (Bell & LaPadula, 1973), RBAC (Sandhu, Coyne, Feinstein, & Youman, 1996), ABAC (Servos & Osborn, 2017), and their combinations (Huang, Nicol, Bobba, & Huh, 2012; Jin, Sandhu, & Krishnan, 2012), applied to their own domains. It will be challenging to create access control policies for AST to work seamlessly with each entity's access control system to reach the targeted goal. In addition to the grave threats of cybersecurity, the models and data about products are also their owners' major concerns about intellectual property protection and business competitivity. All those complex factors have to be taken into account of the access control mechanisms in AST. Basically, it is always a challenging regarding to balancing "need-to-know" and "need-to-share".

**Challenge 5: Scientific Computing Integrity of digital models** -- Scientific computing integrity (SCI) is defined as "the ability to have high confidence that the scientific data that is generated, processed, stored, or transmitted by computers and computer-connected devices has a process, provenance, and correctness that is understood" by DOE ASCR (Advanced Scientific Computing Research) (Peisert, Cybenko, & Jajodia, 2015). Although the concept of SCI is proposed in the context of DOE extreme-scale computing, SCI is also a great challenge to digital engineering. Given the high complexity of engineering workflows across lifecycle phases, across organizational boundaries, and across countries, many models and data produced by different entities in those complex engineering workflows, are going to stream into AST and be reused later. However, the SCI of those models and data could be compromised for many reasons, such as malicious attacks and faults caused by equipment/devices, software, networks, engineers or workers (Huang, 2018). Those risk factors, the complexity of engineering workflows, long and complex supply chains, and long and complex provenance chains of those models and data make ensuring SCI in Digital Engineering extremely difficult, thus posing a significant challenge.

**Challenge 6: Reproducibility and Replicability** – There are different definitions about the relevant concepts of reproducibility, replicability, and generalization. This paper uses a more popular one from NSF (NSF_SBE_AC, 2015), reproducibility is the extent to allow a researcher to independently duplicate the results of a prior study with the same procedures and the same data; replicability is to independently duplicate the results with the same procedures but different data; generalizability is the extent that the results of a study apply to other contexts different from the original one. The journal Nature had a special collection dedicated to the "challenges in irreproducible research" to reveal the "growing alarm about results that cannot be reproduced" (Nature, 2018). Recently, a Nobel laureate retracted her latest paper published in the Science journal for reproducibility issue (BBC, 2020). Earlier, among 1576 researchers surveyed by journal Nature, "More than 70% of researchers have tried and failed to reproduce another scientist's experiments" (Baker, 2016), including engineering. In engineering practice, additional complexity comes from the possible loss of techniques, skills, and know-how knowledge living in human teams.

**Challenge 7: Practical difficulties in producing products' digital counterparts** -- In digital engineering, an enterprise needs to produce not only products (no matter which is hardware, software, or service) and their traditional technical documentation but also the associated digital counterparts including the models for a product, the data supporting the models, as well as associated knowledge. This change is a significant transformation for product producers, thus practically posing some big challenges to enterprises about those digital counterparts. Examples of challenging issues include higher standards on model credibility, repeatability, interpretability, interoperability, intellectual property protection, security, cost, well-trained workforces, and others.

**Challenge 8: Insufficient knowledge in the workforce** – Workforce is essential for the realization of digital engineering. The knowledge and skills required for digital engineering practice are beyond the ones of the traditional engineering workforce and beyond traditional engineering education and training programs. It is a challenge for training a large population of engineers with a varied professional background in the current workforce through on-job training and engineering education programs.

In the emerging digitalized and connected environment, systems engineering is facing many new challenges beyond we discussed above and facing many new research issues. Just list a few: how does an enterprise transform its enterprise culture and policies with sharing engineering artifacts across engineering stages and across multiple organizations? How does the digitalized and connected engineering environment impact human-machine interactions (Handley, 2019), considering both unprecedented rich information and complexity? Furthermore, in this environment, how could teams collaborate (Powell & Pazos, 2017) more efficiently? How does transparency in this environment improve trust (Huang & Nicol, 2013)? What are new risks introduced in digital engineering? How does a better understanding of those risks improve system design (Pinto et al., 2009)? Many interesting and important issues need attention.

## 4. Key Enabling Technologies

We have been working on to identify a set of key enabling technologies for digital engineering. For space limitation, it is impossible to have a comprehensive review of enabling technologies in this paper; here, we briefly discuss several clusters of key enabling technologies. Let us start with AI&ML cluster, which plays a central and fundamental role.

**AI and machine learning cluster**: Enabling to establish a foundation for continuing exploration of automation in digital engineering; enabling digital representation of system of interest; enabling model building through machine learning; enabling intelligent reasoning, control, scheduling, planning, and decision making for digital enterprises. For AI's tremendous impacts to almost every aspect of society, U.S. launched the national AI research and development strategy (US NSTC, 2016, 2019). Earlier, Stanford University's report on one hundred year study on Artificial Intelligence (AI100) (Stanford, 2016) presents a big picture of AI history and future. Machine learning (LeCun, Bengio, & Hinton, 2015) together with Big Data technologies (such as Apache Spark's TensorFlowOnSpark, Apache Hadoop Submarine) will enable building system models with big data gathered from the digitalized and connected system lifecycle. Reinforcement learning (Silver et al., 2018; Sutton & Barto, 2018) (e.g., as achieved by AlphaGo that defeated world #1 player in Go game) will enable to keep improving a system's performance in operations. After a journey from general to domain-focused, AI now again is towards artificial general intelligence (Adams et al., 2012), which is paving the way for innovation of new generation of intelligent engineering systems.

**Ontologies and semantics technologies cluster**: Enabling semantic representation of the general properties of models and their relations; enabling model sharing and integration across boundaries of enterprises, disciplines, and engineering stages; enabling digital representation of enterprise-related concepts and processes. "An ontology is a formal, explicit specification of a shared conceptualization" (Gruber, 1993; Studer, Benjamins, & Fensel, 1998). Ontologies and semantic web are related subfields of AI, focusing on formalizing the semantics and knowledge sharing (Baclawski et al., 2018; Fritzsche et al., 2017). Intensive research on enterprise modeling and enterprise integration has been conducted since 1990s and can be used to support model sharing across boundaries (Chen, Doumeingts, & Vernadat, 2008; Fox & Huang, 2005; Goranson, 2002).

**Provenance modeling cluster**: Enabling to represent and maintain the provenance of engineering artifacts, particularly models; enabling tracing dependency relations among digital engineering artifacts; supporting model reproducibility and replicability; helping with trustworthiness evaluation of digital engineering artifacts. "Provenance information is extremely important for determining the value and integrity of a resource" (Berners-Lee, Hall, Hendler, Shadbolt, & Weitzner, 2006). Buneman, Khanna and Tan (2001) first proposed "Data Provenance" to address where and why issues in complex data workflow. Fox and Huang (2003) proposed "Knowledge Provenance" (KP) to address the problem regarding how to determine the origin and validity of knowledge, by means of modeling and maintaining information sources, information dependencies, as well as trust structures. Research on provenance in eScience and scientific workflow has led to a milestone work "Open Provenance Model" (Moreau et al., 2011); this work was further developed into PROV ontology (W3C Provenance Working Group, 2013), which can support modeling provenance for engineering artifacts.

**Trust management technology cluster**: Enabling to build "Authoritative Source of Truth" with proper trust mechanisms; enabling access control of digital engineering artifacts stored in AST; enabling trust judgment of digital models and artifacts. Many access control paradigms have been developed to meet different needs, such as Bell-LaPadula model (Bell & LaPadula, 1973) based Mandatory Access Control (MAC) (or Multilevel Security) for military and government entities, Role-Based AC (RBAC) (Sandhu et al., 1996) for business world, and more recent Attribute-Based AC (Servos & Osborn, 2017) allowing making more general access policies based on attributes by using standardized policy language XACML (OASIS, 2013). Some frameworks (Huang & Nicol, 2012; Huang et al., 2012; Jin et al., 2012) have been proposed to integrate different models and to meet the new demands of balancing "need-to-know" and "need-to-share", thus achieving the DES goal of delivering "the right data to the right person for the right use at the right time". Research on using AI to build computational trust in distributed environment, e.g. (Cho, Chan, & Adali, 2015; Huang, 2018; Huang & Nicol, 2009; Marsh, 1994) enables to design various trust mechanisms (Huang & Nicol, 2013), from centralized one with standards and certifications to distributed ones based on evidence, distributed attribute certifications, and others. The recent arising Blockchain technology (Wang et al., 2019) enables to verify data integrity in digital engineering, which is an essential mechanism to assure system security.

**High-Performance Computing (HPC), Cloud Computing, and Big Data technologies**: Enabling to store, manage, query, process, mining, analysis, and use a vast number of digital models and engineering artifacts in a manner of scalable, elastic, timely, and ubiquitous access; enabling large scale collaborative research and development across boundaries of disciplines and organizations. Science and engineering have been more and more depending on computing power; extreme-scale computing (ASCR, 2016) has become a core capability for competitive advantage. Digital engineering will depend on computing infrastructures much more than traditional engineering. Cloud computing (Armbrust et al., 2010), HPC, and the associated Big Data technologies (NIST, 2018) comprise the computing infrastructures for digital engineering. In recent years, a trend emerges to combine cloud technologies in scientific HPC for providing highly dynamic and customized computing support (Asch et al., 2018; Keahey & Parashar, 2014). Computing infrastructures for digital engineering should leverage those new developments in computing technologies.

## 5. A Framework of Digital Systems Engineering

Digital engineering is the destination of digital transformation of engineering. Figure 3 illustrates a high-level abstraction of digital engineering transformation, which is characterized by (1) digitalization of engineering artifacts; (2) engineering in a digital and connected environment. There are a variety of engineering artifacts potentially to be digitalized. Digitalized models will play a central role in digital engineering. One engineering process will have inputs of digitalized products and services from other engineering processes in the digital and connected environment and will produce new digitalized products and services. Digital engineering is enabled and facilitated by many new digital technologies including IoT, smart CPS, Big Data technologies, AI, machine learning, digital twin, distributed trust, Blockchain, and others. Among those enabling technologies, knowledge representation & reasoning (KR&R), ontology engineering and semantic web, all of which are branches of AI, play a critical role in digital representation of engineering artifacts particularly digital models. The foundation of KR&R, ontologies, and formal semantics is formal logic. As a matter of fact, in systems engineering, conceptualization has an essential but implicit foundation – logic, however, which has not received much attention; (Dickerson & Mavris, 2010) is one of very few work addressed this critical issue. To reflect this, formal logics together with computing science (which is obviously an essential foundation for digital engineering) are considered as one of the scientific foundations for digital engineering.

## 5.1 Core Concepts

First of all, let us clarify and define a small set of core concepts. Sometimes, terms "digitization" and "digitalization" are used as equivalently, but they have different meanings. According to Gartner IT Glossary, "Digitization is the process of changing from analog to digital form" (www.gartner.com/it-glossary/digitization/); "Digitalization is the use of digital technologies to change a business model and provide new revenue and value-producing opportunities; it is the process of moving to a digital business." (www.gartner.com/it-glossary/digitalization/). Gartner's definition of digitalization is from business perspective. Technically, in order to enable using digital technologies, a digitalized item should be not only computerized but also in a standard form and annotated with necessary metadata to enable machines of different types to access and operate automatically. From this perspective, a digitized item can be in a preliminary digital form and is not fully digitalized. For example, a pure image file in an ad hoc format is a digitized photo but not digitalized; a digital photo in standardized format with metadata about the technical parameters used, the device, the time, and the location where the picture was taken (as the one taken by a digital camera) is digitalized. In another example, in the case that a physical book was scanned into a computer or directly typed as text in a computer, we call it "digitized"; in the case that a book was produced as eBook with attached metadata to allow properly displayed by different reading software and protected by digital right management software, we call it "digitalized". Generally, "digitalize" is beyond "digitize".

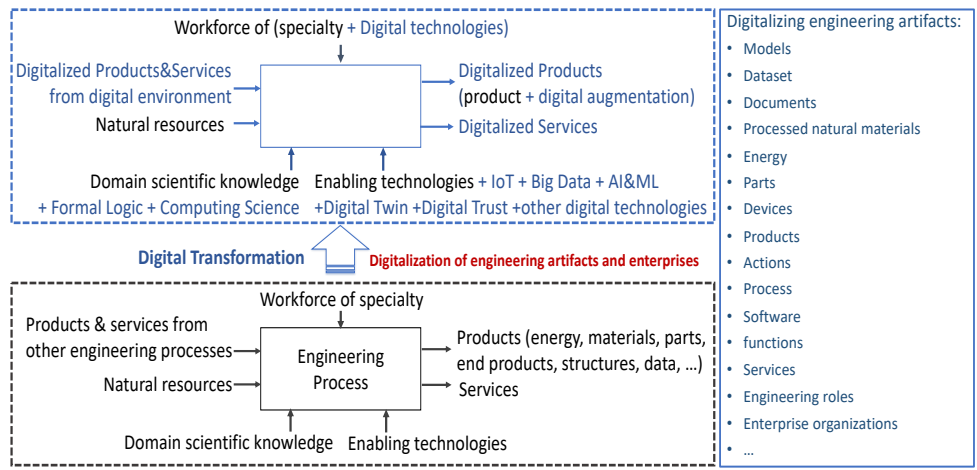

Figure 3. Digital transformation of engineering

### 5.1.1 Digitalization

To clearly characterize the difference discussed above, we define "digitalize" as follows.

"**Digitalize**" is to (1) digitally represent an item or thing in a standard form with well-defined semantics to make it universally accessible by different types of machines; (2) assign and maintain an unique identifier to the item or thing; (3) create necessary metadata in a standard form with well-defined semantics, to enable the use of digital technologies to manipulate and operate that item or thing automatically; (4) uniquely associate the unique identifier and the metadata with the item or thing.

The unique association of metadata and a unique identifier with an artifact can be achieved by using RDID, digital signature, and Blockchain.

There could be different **degrees of digitalization**, dependent on the degree of fidelity of the digital representation, the depth of semantic representation, the compatibility of the representing form, and the richness of metadata, which allow a different extent of digital operations and interoperations by machines of different types automatically. A digitized thing can be in the simplest digital form, while a fully digitalized thing with a high degree of digitalization is in the high end of digital form. The loosely used term "**digital xxx**" could be across a broad spectrum from simple to highly digitalized. By leveraging emerging innovative digital technologies, a digitalized engineering artifact can gain many new advantages such as enhanced or improved performance, fast and agile integration in new systems, enabled traceability and accountability, sharing across the boundaries of organizations and lifecycle activities, and many others.

### 5.1.2 Unique Identification

**Unique identification** is a critical part of digitalization because it is a necessary component for traceability and accountability. (Note that the "**traceability**" we address here is not limited to requirement traceability in systems engineering; instead, it is a general term about tracing information flow, material flow, chain of causality, as well as chains of faults.) Barcode and RFID are good examples for better understanding the profound impacts of unique identification.

### 5.1.3 Digitalized Artifacts and Digital Augmentation

Based on the definition of term "digitalize", a **digitalized artifact** consists of the artifact and its **digital augmentation**, which consists of (1) digital representation, by which the artifact is represented in a standard form with well-defined semantics thus accessible by different types of machines; (2) an identifier, which is uniquely associated with the artifact to enable traceability and accountability; (3) associated metadata, in a standard form with well-defined semantics, to enable the use of digital technologies to manipulate or operate the artifact.

An artifact could be either a digital object such as a model, a dataset, a document, a picture, and others, or a physical object such as a physical product or a part. For a digital object, the digital representation of it is the object itself. For a physical object, (1) the digital representation of that physical object could be as complex as the digital twin of it, or as simple as just a picture or text description to characterize it; (2) the identifier could be the barcode or the id code of the RFID tag attached to that physical object; (3) the associated metadata covers the properties of the object and its digital representation.

Among various engineering artifacts, models are a particularly important one for digital engineering. In the DoD DES, models will play a crucial role because "digital representation" of systems of interest is a central theme. We believe digitalized models will enable to leverage the power of innovative digital technologies maximally. In the following, on top of the earlier definition of digitalized artifact, we further conceptualize "digitalized model" with different aspects of metadata.

### 5.1.4 Digitalized Models

A **digitalized model** is composed of a model and its digital augmentation, which consists of (1) the digital representation of the model, which is actually the model itself, if the model is in a digital form; (2) a unique identifier associated with the model to enable traceability and accountability; (3) associated metadata, in a standard form with well-defined semantics, to enable to use or interact with the model by engineering systems using digital technologies. More specifically, the metadata about the properties of the model include possibly:

- Generic attributes, such as model name, version, date of creation, type of the model, and others;
- Description of inputs, outputs, and parameters;
- Provenance of the model, such as, who created the model, when the model was created, why the model was created (purpose), where the model was developed and tested (computing environments), what the model depends on (dependence relations to other models, datasets, documents, and others ), the revision history of the model, and others;
- Model utilization guide about how and where the model can be viewed, executed, or used.
- A set of security properties about the model such as check sum (hash code), security label, digital signatures, various certificates, and others.
- A set of machine-processible access control policies, to enable an external guard system to enforce the specified policies, or a self-contained access control software module to protect the model.

The models that can be digitalized include all types of models, no matter they are mathematical models in print or in a digital form, logic models (a subset of mathematical models, e.g., in First Order Logic, modal logic, temporal logic, and others), an executable model represented with a programing language, an engineering design, a conceptual graphic model (e.g., flowcharts), or others.

The term "standard" means to follow precisely defined syntax and semantics, which are commonly shared by a community. Therefore, a "standard" form could be a form that is complying with a set of officially issued standards or with ontologies commonly used in a community.

An officially issued standard represents a centralized approach, which is highly efficient at least in the short term, but maybe a sub-optimized solution representing a local optimum from long term view. Ontologies represent a decentralized or distributed crowdsourcing evolutionary and fine-grained level of standardization approach; a concept may be formalized and published as several ontologies by several participants, but the one that is mostly reused by others becomes de facto "standard". A concept is typically defined on top of other concepts, so the evolution of ontologies represents a fine-grained "standardization" process. For short term, the use of ontologies is not efficient as a standard does, as ontology mapping is usually needed and sometimes can be complicated and inefficient; for the long term, the use of ontologies leverages collective intelligence, incentivizes innovation, and allows evolutionary revision in micro-level of a "standard".

When a set of related ontologies become mature, they can be adopted and issued as an official standard by an authority for a community.

To build the digital augmentations for models, we need to investigate and understand the relevant properties of different types of models and their logical relations. We need to use knowledge representation to explicitly model and represent those properties and relations. We call this type of models as "**model of models**". The study of "model of models" is a foundation for developing ontologies used to express the digital augmentations for models.

## 5.2 Overarching Goal and Focusing Areas

The US DoD digital engineering strategy presents a profound vision for the emerging digital engineering, specific to the engineering practice of the DoD enterprise; the strategy is also inspiring to the development of digital systems engineering as an academic research field to develop corresponding scientific knowledge and technology to support the proposed digital engineering transformation. In this section, we attempt to draft a general framework for digital systems engineering in the global context of digital transformation, Industry 4.0, and Data Science. The development of digital systems engineering will support the implementation of the DoD DES as well as digital engineering in general with knowledge, methodologies, technologies, as well as training and education for the workforce.

The core of digital systems engineering is **digitalization**. The broader overarching goal of digital systems engineering is to develop the principles, theories, methodologies, methods, models, and technologies for the digitalization of engineering and for systems engineering in the digitalized and connected engineering and operating environments.

The immediate targets of digital systems engineering are the digitalization of engineering artifacts, information and model sharing, and the associated issues of digital trust, big data, automatic machine-processing, and machine learning arising from the digitalized and connected environment.

Digital systems engineering first needs to digitalize a variety of engineering artifacts, such as models, data, materials, products, services, processes, enterprise, and others. As part of digitalization, **unique identification** plays a critical role to enable accountability and traceability (for tracing information flow, material flow, faulty chain, supply chain, and others). A central task is to develop digital augmentation for each engineering artifact with well-defined semantics, thus enabling the use of digital technologies to manipulate, operate, or interact with those engineering artifacts.

If digitalization is a mean to enable rich information for digital engineering, we also need to effectively and efficiently deal with those big data. In a digitalized and connected environment, every phase of systems engineering lifecycle will face unprecedented rich information; it is a great challenge regarding how to leverage those big data, which is another focusing theme of digital systems engineering. On this matter, two research issues need immediate attention. First, how should the big data be handled in digital engineering? How can we leverage Data Science to gain insights from those big data in the domain of digital engineering? Secondly, given the distributed nature of data in the digitalized and connected environment, the trustworthiness of digitalized engineering artifacts (including models) is a critical issue. What digital trust mechanisms will be needed?

To address the above research issues, digital systems engineering needs to integrate and leverage digital technologies such as Big Data technologies (including cloud computing), Data Science, ML, AI, semantics technologies, as well as digital mechanisms of security and trust developed in cybersecurity, Blockchain, and computational trust communities.

Regarding disciplinary relations, "Digital Engineering" is generally engineering practice in the digitalized and connected environment to leverage digital technologies. Digital systems engineering is an academic research field to develop corresponding scientific knowledge and technology to support digital engineering practice. From the perspective of digital transformation, digital systems engineering is the digitalization of systems engineering; from the perspective of data science, digital systems engineering is an extension of data science applied to systems engineering. Digital systems engineering is a new development of systems engineering by leveraging digital technologies; as a subfield of SE, digital systems engineering is guided by systems thinking, systems approach, and SE principles and methodologies. Within the domain of SE, digital systems engineering is highly relevant to an active research area -- model-based systems engineering (Estefan, 2007; Madni & Sievers, 2018). According to INCOSE's definition, "Model-based systems engineering (MBSE) is the formalized application of modeling to support system requirements, design, analysis, verification and validation activities beginning in the conceptual design phase and continuing throughout development and later life cycle phases" (INCOSE, 2007). The central goal of MBSE is to transform the traditional document-centric approaches to systems engineering into model-centric approaches, thus overcoming the deficiencies of the former (INCOSE, 2007; Madni and Sievers, 2018). As stated earlier in this section, the core of digital systems engineering is digitalization. Digital systems engineering focuses on digitalization (including unique identification) of engineering artifacts, as well as associated big data and distributed trust issues in digital engineering. Digitalization enables machine-processable (understandable) digitalized artifacts and digitalized engineering processes, thus enabling adopting, infusing, or integrating new digital technologies rapidly and smoothly; unique identification (as a part of digitalization) enables information traceability and accountability in systems lifecycle. Digital systems engineering works together with MBSE to support digital engineering. The relation stated above can be illustrated in figure 4. Digital systems engineering also supports System of Systems Engineering (C. Keating et al., 2003) and Mission Engineering (Gold, 2016; Sousa-Poza, 2015) with enriched information for complex system governance (C. B. Keating & Katina, 2019) and systems coordination and planning.

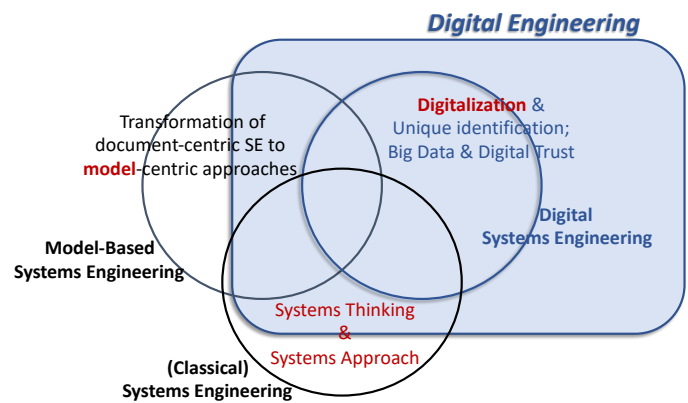

Fig. 4. Relations between digital systems engineering and MBSE as well as classical SE.

From the perspective to realize the digital engineering vision, the knowledge and research areas of digital systems engineering can be organized in four levels: vision, strategy, action, and foundation, as illustrated in figure 5.

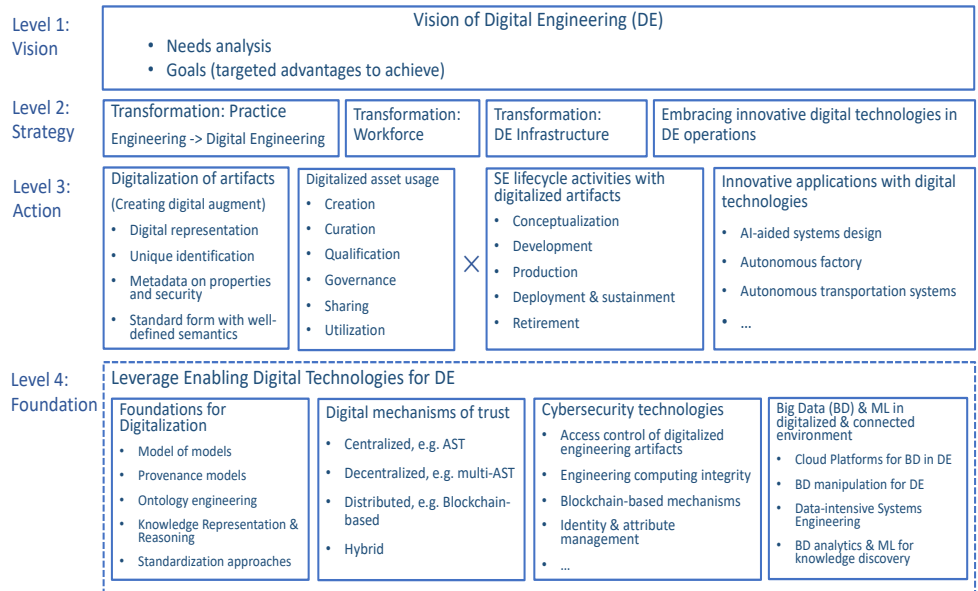

Figure 5. Knowledge and Research Areas of Digital Systems Engineering

### 5.2.1 Vision Level

**Vision of Digital Engineering**: Digital systems engineering aims to develop means towards digital engineering; therefore, it is crucial to have a clear vision of digital engineering. The US DoD DES is an insightful vision but specific to DoD enterprise. It is important to explore the vision of digital engineering in general, i.e. engineering in the coming digital age. Research areas for this purpose include:

- Needs analysis
    - Features of fast-changing engineering and business environment
    - Deficiencies of current engineering practice
    - Conceptualization (envisioned future digital engineering)
    - Gap analysis
    - Enabling technologies
    - Feasibility study
- Goals of digital engineering (the targeted advantages to achieve)

### 5.2.2 Strategy Level

At strategy-level, we identify four strategic moves.

**Transformation of engineering practice**: In digital engineering transformation, the central task is the digitalization of engineering artifacts, which will be discussed later.

**Transformation of education and workforce development**: Human is the essential force to advance engineering; therefore, workforce development and education transformation towards digital engineering are necessary. There are two related aspects:
- Education and training with knowledge and skills for digital engineering
- AI-powered innovative learning & training systems in digital engineering environment

**Transformation of engineering infrastructure**: To facilitate digital engineering, it is also necessary to develop digital engineering infrastructure for meeting the demands from Big Data, security, and distributed nature of a digital & connected environment.
- Cloud-based platforms for digital engineering
- Digital mechanisms of trust and security for digital engineering

**Embracing innovative digital technologies in digital engineering operations**: After realizing digital engineering, an enterprise is able to take full advantages of digitalization by fast adopting, interacting, and/or integrating with emerging innovative digital technologies, thus achieving fast design, delivery, and sustainment of agile intelligent and complex systems in fast changing environment.

### 5.2.3 Action Level

At action-level, there are many research areas, roughly organized in four groups.

**Digitalization of engineering artifacts**: Towards digital engineering, it is a critical step to investigate how to digitalize engineering artifacts, potentially including various models (formal or informal, numeric or logical, abstract or physical), datasets, various documents, bills of materials, processed natural materials, energy, parts, devices, products, actions, process, software, functions, services, engineering roles, enterprise organizations, and others. In the digitalization, **unique identification** plays a critical role to enable accountability and information traceability (for tracing information flow, material flow, faulty chain, supply chain, and others). To digitalize engineering artifacts, a digital augmentation will be developed for each artifact, as discussed in subsection 5.1.

**Operations of digitalized engineering artifacts**: Once the means to digitalize engineering artifacts is created, digital engineering practice will need to explore the technologies regarding the following operational aspects of digital engineering artifacts:
- Creation (manual and automatic approaches)
- Curation (store, organization, query, retrieval, change, upgrade, …)
- Qualification (consistency, validity, completeness, usability, accessibility, …)
- Governance (policies of access, sharing, security, intellectual properties, …)
- Sharing (information flow across organizations and lifecycle activities)
- Utilization

Each type of digitalized engineering artifacts has different features and needs further studies to look into them individually. Given the critical role of models in digital engineering, digitalized models need immediate attention. Some interesting research areas include but not limited to:
- Automatic generation of digital augmentations for models by machine in an Integrated Development Environment for system design
- Digitalized model creation in a digital environment with a large number of relevant digitalized artifacts

- Digitalized model creation by integrating a set of existing digitalized models for its components
- Digitalized model creation by big data analytics and machine learning
- Digitalized model verification and validation
- Digitalized model curation
- Digitalized model sharing
    - Sharing across engineering stages or lifecycle phases
    - Sharing across boundaries of disciplines and organizations
    - Digitalized model update and propagation
- Digitalized model repeatability and reusability
- Digitalized model interpretability
- Digitalized model usability
- Digitalized model interoperability
- Digitalized model trustworthiness evaluation
- Digitalized model access control
- Digitalized model and intellectual property protection
- Digitalized model security and risk analysis

**Systems engineering lifecycle activities with digitalized engineering artifacts**:
Every phase of SE lifecycle will involve operations of digitalized engineering artifacts discussed above and will have and need to investigate how to leverage unprecedented information from the digitalized engineering artifacts of upstream lifecycle phases, from historical observations of the same or similar work in downstream of lifecycle, and from a digital connected environment. Research issues appear in every combination of each digital artifact operation type and each SE lifecycle activity, similar to figure 6, which illustrates digitalized model operations vs systems engineering lifecycle phases in a digitalized environment.

|  | **Concept Stage** | **Development Stage** | **Production Stage** | **Utilization/support Stage** | **Retirement Stage** |
|---|---|---|---|---|---|
| Model creation. | Inputs:<br>- Digital artifacts (DAs) in operating environment;<br>- Relevant data and models from downstream stages. | Inputs:<br>- Digital artifacts from both upstream and downstream;<br>- Digital artifacts of external systems | Inputs:<br>- Digital artifacts from both upstream and downstream;<br>- and from production environment. | Inputs:<br>- Digital artifacts from both upstream and downstream;<br>- and from operating environment. | Inputs:<br>- Digital artifacts from upstream;<br>- and from natural environment. |
| Model learning | Apply AI&ML for model building from big data coming from upstream and downstream engineering stages and environment. | | | | |
| Model integration | Interaction between digital models for both SysCon and systems in operating environment. | Interaction between digital models for both system components and external systems. | Interaction between digital models for both the system and systems in production environment. | Interaction between digital models for both the system and external systems in operating environment. | Interaction between digital models for the sys component and DAs in natural environment. |
| Model curation | Create model of models; maintain metadata for models; maintain model provenance; model update and propagation to downstream. | | | | |
| Model sharing & use | Across lifecycle activities,; across the boundaries of disciplines and organizations | | | | |
| Model Trustworthiness | Centralized standardization; decentralized standardization and mappings; distributed evolutionary fine-grained convergence; model trustworthiness; model repeatability; Access Control; digital artifact intellectual property protection, … | | | | |

Figure 6. Operations of models in systems engineering lifecycle within a digitalized and connected environment.

**Innovative applications with digital technologies**: Digitalized engineering practice will enable to gain the advantages of many digital technologies and develop innovative applications. Number a few as follows.
- AI-aided systems design
- Autonomous factory
- Autonomous transportation systems
- …

There are many potential innovative applications. We briefly discuss systems design with digitalized engineering artifacts. There could be several levels of automation. At a basic level, a digitalized model can be just an engineering diagram for human users, wrapped with metadata about the diagram; following the information given in its metadata, the diagram can be displayed with the required software. At a middle level of formalism, a digitalized model can be an executable formal model; following the associated metadata, the formal model can be executed in an environment such as a container (a modern approach of virtualization in computing); the origin and the evolution history of the model can be traced; the dependency relations of the model can be traced. At an advanced level, in addition to what stated above, some examples of the capabilities include: (a) using the metadata, a digitalized model can be verified and validated by machines, e.g., using a model checker or a theorem prover to prove the validity of a logic model. (b) Some types of faults in a model can also be possibly found by machine automatically. (c) Based on the properties of a set of digitalized models, they are integrated by human modelers to construct a model for a system; At highest level, by using AI, a machine can autonomously construct a model on top of available digitalized artifacts; a machine can autonomously improve a digitalized system's structure and behaviors.

It is a strategic goal of digital engineering to leverage innovative digital technologies, such as IoT, CPS, Big Data technologies, Data Science, AI (including ML, KR&R, ontologies, Semantic Web), Augmented Reality (AR), Virtual Reality (VR), digital twin and digital thread, 3D printing, modeling and simulation (M&S), cybersecurity, distributed trust, Blockchain, and others. In the history of engineering, CAD has largely digitalized engineering drawing and the detail physical design. Combining with VR&AR and M&S, CAD is moving towards digital twins. No doubt there will be many innovative applications by combining digital engineering with other emerging digital technologies.

### 5.2.4 Foundation Level

At foundation-level, we identify several groups of foundational research areas and enabling digital technologies for digital engineering.

**Foundation for digitalization**: The main foundation of digitalization is from AI, more specifically, knowledge representation & reasoning (KR&R) and semantics technology developed from ontology engineering and Semantic Web communities. Foundational research areas include:
- Model of models, to categorize models and to create logic model for representing each category of models' properties and their relations.
- Provenance modeling, to create logic models for representing the origin and the dependency relations of engineering artifacts.
- Ontologies for expressing the digital augmentations with well-defined semantics, thus enabling or enhancing model integration, model repeatability and model reusability, model curation and sharing across engineering stages and across the boundaries of disciplines, teams, and organizations.
- Standardization in digitalization
  - Centralized standardization
  - Decentralized standardization and mappings
  - Distributed fine-grained evolutionary convergence with ontologies

**Digital mechanisms of trust**: Given the distributed nature of digitalized and connected engineering environment, the trustworthiness of digitalized engineering artifacts (particularly models) is a critical issue. What digital mechanisms of trust will be needed? The US DoD's "Authoritative Source of Trust" (AST) is a centralized solution; what specific mechanisms of trust need to be used in the AST? How other distributed digital trust mechanisms can be incorporated in AST? Interesting mechanism types include:
- Centralized mechanisms, e.g. US DOD's AST
- Decentralized mechanisms, e.g. multiple ASTs in different domains with certain structures such as hierarchical and mesh as for PKIs (Huang & Nicol, 2009).
- Distributed mechanisms, e.g. using digital signature, distributed key certification and attribute certification, evidence-based trust, and blockchain.
- Hybrid mechanisms.

**Cybersecurity technologies cluster**: It is of paramount importance to ensure the security of digital engineering and digital enterprises. Worrying failure in cybersecurity can be a factor to block digital engineering transformation. There are many interesting research areas to support the security of digital engineering, just numbering a few,
- Access control of digitalized engineering artifacts
- Engineering computing integrity in a distributed digital engineering environment
- Identity and attributes management in digital engineering
- Blockchains-based distributed mechanisms of trust and security
- Intrusion detection in a distributed digital engineering environment
- …

Cybersecurity in digital engineering will be discussed in another paper.

**Big Data and Machine Learning cluster**: In the digitalized and connected environment, given those unprecedented big data characterized by volume, velocity, variety, veracity, and views (Huang, 2018), how do we design and build trustworthy AI systems and ML models and algorithms for knowledge discovery from big data to ensure reliable performance, explainability, safety, security, resilience, scientific computing integrity in digital engineering? Some foundational research areas in this cluster include:
- Cloud platforms for big data in digital engineering
- Big data manipulation for digital engineering
- Big data analytics and machine learning for knowledge discovery from observations of the system and its environment in digital engineering
- Data-intensive Systems Engineering (SE combined with insights and findings discovered from data through big data analytics and machine learning)

We have discussed some knowledge and research areas broadly in four levels from vision, strategy, action, to foundation, as illustrated in figure 5. No doubt, many other interesting and important areas were not reflected in the figure for limited space. Some examples are:
- Human-machine interaction in a digitalized connected environment
- Team collaboration in a digitalized connected environment
- Parts recycle & reuse evaluation in a digitalized connected environment
- Environment impact analysis in a digitalized connected environment.

## 6. Concluding Remarks

"Digital engineering will require new methods, processes, and tools" (US DoD, 2018). To support the digital engineering strategy, digital systems engineering is emerging as an academic field, which aims at developing theory, methods, models, and tools to support digital engineering practice in the emerging digitalize and connected environment. For that end, in this paper, we (1) analyzed the transformation from traditional engineering to digital engineering (figure 3); (2) clarified and defined a small set of core concepts including digitalization, digitalized artifacts, digital augmentation, and digitalized models; (3) presented a big picture of digital systems engineering in four levels: vision, strategy, action, and foundation (figure 5), and discussed each of identified main areas of research issues. A critical task towards digital engineering is to digitalize engineering artifacts, including models, datasets, products, functions, services, and SE processes. Digitalization enables universally machine-processable (understandable) digital engineering artifacts and processes, thus enabling rapid infusing and leveraging innovative digital technologies; as a part of digitalization, unique identification plays a critical role to enable information traceability and accountability in systems lifecycle. In the digital engineering information flow, provenance also plays critical role in enabling tracing the dependency relations among engineering artifacts and improving model reproducibility and replicability.

This paper presented our vision on digital systems engineering, and much work is ahead in that direction. Numbering a few, we will explore the ontological approach to digitalizing engineering artifacts with higher priority, developing model of models for digitalizing models, and developing provenance representation and reasoning models. We will research the distributed digital trust mechanisms for digital identity and attribute management and for digital engineering artifacts sharing in more general engineering collaboration in today's business environment. We will explore data-intensive systems engineering approach by researching trustworthy AI systems and ML models and algorithms for knowledge discovery from big data to ensure reliable performance, explainability, safety, security, resilience, scientific computing integrity in digital engineering.

The development of digital systems engineering will support the implementation of DoD DES as well as digital engineering in general with the needed knowledge, methods, technologies, as well as training and education for the workforces.

## Acknowledgment

This work is partially supported by the National Science Foundation under grant CNS-1828593. This paper is a revision and further development of our earlier white paper on digital systems engineering (Huang et al., 2019).